# Temperature-Modulated Photomechanical Actuation of Photoactive Liquid Crystal Elastomers

Zhengxuan Wei, Ruobing Bai*

Department of Mechanical and Industrial Engineering, College of Engineering, Northeastern University, Boston, MA, 02115, USA
*Corresponding author: ru.bai@northeastern.edu

**Abstract**

Photoactive liquid crystal elastomers are polymer networks of liquid crystal mesogens embedded with chromophores like azobenzene. They undergo large deformation when illuminated by light of a certain wavelength through photochemical reaction, inspiring exciting new applications. However, despite the recent progresses in both the experiment and theory of these materials, the fundamental understanding of the temperature effect on their photomechanical actuation through various molecular-to-mesoscale processes have remained largely unexplored. This paper constructs a theoretical model to investigate this temperature-modulated photomechanical actuation, by integrating different temperature-dependent processes into a continuum framework. The model studies a special working condition where the material is subjected to a uniaxial tensile load, a prescribed temperature, and a polarized light illumination. We explore the free energy landscape of the system and the uniaxial stress-stretch responses under various conditions. We exploit the coupling between individual controls of temperature and light in a single photomechanical actuation for several working scenarios, including the temperature-modulated photomechanical snap-through instability, specific work, and blocking stress. We study the effect of the temperature-dependent backward isomerization of chromophores on the photomechanical actuation. These results are hoped to motivate future fundamental studies and new applications of various photomechanical material systems.

## 1. Introduction

Modern challenges faced by human society have motivated a new paradigm of internet of materials with inherent physical intelligence [1, 2]: materials can operate in complex working conditions, capable of sensing, actuation, and analysis with self-sustainable power supply, without a need for central computing or local energy storage. Photomechanical actuation, a direct energy conversion from light to mechanical motion through photochemistry, is particularly desirable for this new paradigm. This actuation is widely available in nearly all materials: crystals [3, 4], glassy polymers [5, 6], semicrystalline polymers [7], and elastomers [8, 9]. The actuation can be highly reversible and repeatable [7, 10], with fast speed through diffusionless martensitic-like phase transformation [10, 11]. The actuation results from the photochemical reaction of chromophores such as azobenzenes, providing a remarkable work output with density of $\sim 10^8$ J/m$^3$, thermodynamic efficiency of $\sim 10\%$, and reaction time of picoseconds at the molecular level [12]. These attractive features have inspired exciting new applications including light motors [13], shape morphing structures [14, 15], and light robots with swimming [16], crawling [17], and rolling capabilities [18]. Accompanying new applications are the rapid developments of novel photoactive molecules [19-21], geometries and structures [5, 17, 22], and useful theories for photomechanical responses under complex loading conditions [8, 10, 11, 23-27].

With the rapid development of photomechanical actuation in all aspects, it is foreseeable that photomechanical material systems will soon be pushed to applications in various and even extreme working conditions. Among other factors, temperature is particularly important to the photomechanical responses of all existing material systems. Take a major class of photomechanical material—photoactive liquid crystal elastomer (LCE) as an example. The material consists of a rubbery polymer network with liquid crystal mesogens linked to its polymer chains and photoactive chromophores such as azobenzenes embedded in or linked to the network. During a photomechanical actuation, temperature critically affects processes and properties including the nematic-isotropic phase transformation of mesogens [28], the entropic elastic modulus of the polymer network [28, 29], and the reaction kinetics of the backward *cis-trans* isomerization of azobenzene [30]. Each of these temperature-dependent aspects has been well characterized and modeled for decades in thermo-responsive liquid crystals, LCEs, and photoactive chromophores. However, their coupling effects on the actuation of a photoactive LCE have been poorly explored so far, despite some preliminary experimental evidence [31, 32]. This lack of fundamental understanding clearly hinders the further development of photomechanical materials and the expansion of their large-scale applications.

We have recently built theoretical and numerical models for temperature-dependent photomechanical coupling in photoactive molecular crystals [11] and semicrystalline polymers [10] , with good qualitative agreements to experimental observations [7, 33]. We have also recently established a continuum framework that incorporates various micro-to-mesoscale processes to investigate photomechanical coupling in photoactive LCEs [27]. These fundamental progresses readily set the stage for the further research of temperature-modulated photomechanical actuation of photoactive LCEs, which is the goal of the current work.

This paper constructs a theoretical model to investigate temperature-modulated photomechanical actuation of a photoactive LCE for the first time. We start in Section 2 by integrating different temperature-dependent processes into a continuum framework of photoactive LCE. We use the model to study a special working condition, where an LCE is subjected to a uniaxial tensile load, a prescribed temperature, and a polarized light illumination, with the tensile stress and light polarization both parallel with the nematic director. We investigate the free energy landscape of the system with various light intensities, stress magnitudes, and temperatures, together with the uniaxial stress-stretch responses of the LCE under different conditions in Section 3. We exploit the coupling between individual controls of temperature and light in a single photomechanical actuation for several working scenarios in Section 4, including the temperature-modulated photomechanical snap-through instability, specific work, and blocking stress. Section 5 further studies the effect of the temperature-dependent backward isomerization of chromophores on the photomechanical actuation. Finally, discussions on future research and application of photomechanical actuation are provided in Section 6.

## 2. Microscopic picture and continuum framework

The general microscopic process in the photomechanical actuation of a photoactive LCE is illustrated in **Fig. 1a**. We focus on a main-chain LCE, where the liquid crystal mesogens are linked in the backbone of the polymer network. At room temperature, the material stays in the *nematic* phase, where the mesogens form a directional order due to their molecular interaction or steric effect. The average aligned direction is defined by the nematic director **n**. The degree of alignment is defined by the nematic order parameter $Q$ ($Q = 0$ for the isotropic phase and $Q = 1$ for perfect alignment). The LCE is embedded with azobenzene chromophores, which stay in the rod-like *trans* state without any light illumination, and isomerize to the bent-shaped *cis* state when illuminated with a light of certain wavelength. This shape change of azobenzene reduces the mesoscale nematic order in the LCE, and subsequently induces a macroscopic deformation of the polymer network. The thermal-induced backward isomerization of azobenzene from *cis* to *trans* further makes the deformation reversible.

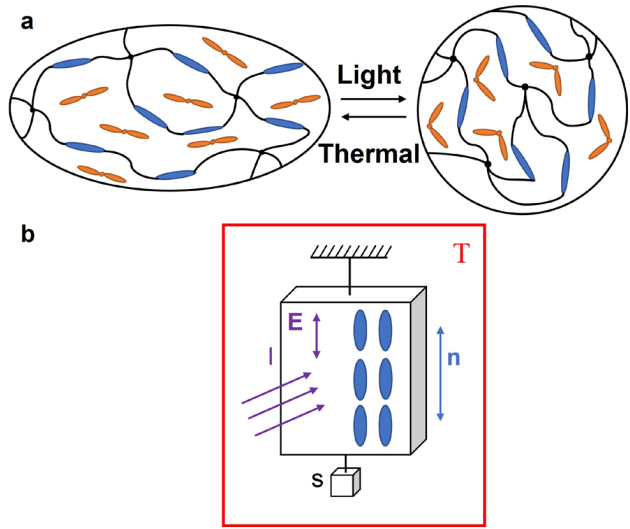

**Fig. 1. (a)** Microscopic process in a main-chain photoactive LCE during photomechanical actuation. Blue rods: non-photoactive liquid crystal mesogens. Black lines: polymer chains. Black dots: crosslinks. Orange rods: azobenzene chromophores in the *trans* (rod-shape) or *cis* (bent-shape) state. **(b)** A thin sheet of monodomain LCE is subjected to a uniaxial nominal stress **s**, a linearly polarized light illumination with intensity $I$, and a prescribed temperature $T$. Both the nematic director (**n**) and the polarization (**E**) are parallel with the uniaxial load. The thickness of the sheet is much smaller than the decay length of the light.

In this paper, we study a specific case of a *monodomain* photoactive LCE: the nematic director **n** and the order parameter $Q$ are uniform in the material. A thin sheet made of such an LCE is simultaneously subjected to a uniaxial nominal stress **s**, a linearly polarized light with intensity $I$ illuminating on the entire surface, and a prescribed temperature $T$ (**Fig. 1b**). Both the nematic director and the polarization are parallel with the uniaxial load. The thickness of the sheet is much smaller than the decay length of the light, such that the light intensity is assumed homogeneous in the sheet. During the actuation, we assume there is no nematic director reorientation, which is reasonable for a monodomain LCE synthesized in the nematic phase with a "memory" effect in their polymer chains [27].

We construct a continuum framework of the above system by adapting the theoretical model we recently developed [27] based on the pioneering work of Corbet and Warner [25, 26]. In particular, the adapted model in the current work considers the temperature effect on various processes during the photomechanical actuation under both light and mechanical load, which was not investigated previously.

We choose the high-temperature, isotropic LCE as the reference state. In the current state shown in Fig. 1b, since the LCE is monodomain, with a homogeneous field of light intensity and without any reorientation of the nematic director, the stress and stretch fields are assumed to be homogeneous. Because of symmetry, the three principal stretches are expressed as $\lambda$, $1/\sqrt{\lambda}$, and $1/\sqrt{\lambda}$, where $\lambda$ is the stretch along the uniaxial load, and the LCE is assumed incompressible. Fig. 1b forms a driven system, with light continuously providing photon energy to drive it away from thermodynamic equilibrium. For such a system, we focus on its *photostationary* state, where the light-driven forward reaction and the thermal-driven backward reaction of the azobenzene, as well as all the other physical processes, reach equilibrium. The total Helmholtz free energy density considering the photoreaction equilibrium is expressed as

$$W(\lambda, Q) = W_e(\lambda, Q) + W_{lc}(Q) - s\lambda, \tag{1}$$

where $W_e$ is the entropic elastic energy of the polymer network [29, 34], $W_{lc}$ is the Maier-Saupe free energy of the nematic mixture [25, 26, 35], and $-s\lambda$ is the potential energy corresponding to the stress **s**. The expressions of the two free energies are

$$W_e = \frac{NkT}{2}[\frac{\lambda^2}{(1+2Q)} + \frac{2}{\lambda(1-Q)} + \log(1-Q)^2(1+2Q)], \tag{2}$$

$$W_{lc}(Q) = N_n kT(1 - c(Q))[g^{-1}(Q)Q - \log Z(Q) - \frac{1}{2}(1 - c(Q)\frac{J}{kT}Q^2)], \tag{3}$$

where

$$g(x) = -\frac{1}{2} - \frac{1}{2x} + \frac{1}{2x}\sqrt{\frac{3x}{2}} \frac{\exp(3x/2)}{\int_0^{\sqrt{3x/2}} \exp(y^2) dy}, \tag{4}$$

$$Z(Q) = \frac{\exp[g^{-1}(Q)]}{1 + g^{-1}(Q)(1 + 2Q)}. \tag{5}$$

In Equations (2)-(5), $N$ is the number of polymer chains per unit volume, $kT$ is the temperature in the unit of energy, $N_n$ is the number of mesogens per unit volume, $c(Q)$ is the fraction of *cis*-mesogens, and $J$ denotes the interaction between mesogens in the unit of energy. In the photoreaction equilibrium, $c(Q)$ is expressed as

$$c(Q) = f\frac{E^2\Gamma\tau(1+2Q)}{3 + E^2\Gamma\tau(1+2Q)}. \tag{6}$$

The constant $f$ denotes the fraction of azobenzene (both *trans* and *cis*) in the total mesogens (both non-photoactive mesogens and azobenzenes). $I = E^2\Gamma\tau$ denotes the dimensionless light intensity, composed by the electric field $E$, a constant $\Gamma$ as a prefactor of light-driven forward reaction kinetics, and the relaxation time of the *cis* chromophore $\tau$. Since the backward reaction from *cis* to *trans* is thermally driven, $\tau$ is temperature dependent. In the following, we will first assume $\tau$ to be constant to investigate the temperature effect through other processes, and later discuss the temperature effect through thermal relaxation in Section 5.

We will minimize the free energy in Equation (1) to investigate the photostationary state of the system. To reduce the number of parameters in our model, we introduce the following dimensionless groups as listed in **Table 1**. To effectively manifest all the temperature effects in different processes, we choose a constant reference temperature $T_{ni}$, the nematic-isotropic transition temperature of the LCE, in forming these dimensionless groups.

Finally, we provide the following representative values of the material and experimental parameters that form the dimensionless groups, to connect the theoretical results from the current work to real experiments. The nematic-isotropic transition temperature of most existing LCEs, $T_{ni}$, typically lies around 60-100 °C [28, 36]. The dimensionless group $\hat{N} = N / N_n = 0.05$ indicates 20 mesogens per polymer chain in the current system. The stress is normalized by a representative shear modulus $NkT_{ni}$, which for most LCEs is on the order of 1 MPa [37]. The light intensity needed for inducing a large uniaxial deformation (with a change of stretch ~0.1-1) is typically on the order of 10-100 mW/cm$^2$ [17, 18, 32].

**Table 1. Dimensionless groups in the model**

| Normalized term | Dimensionless group | Value |
|---|---|---|
| Number density of polymer chains | $\hat{N} = N / N_n$ | 0.05 |
| Free energy | $\hat{W} = W / \left( N_n k T_{ni} \right)$ | N/A |
| Temperature | $\hat{T} = T / T_{ni}$ | Variable |
| Stress | $\hat{s} = s / \left( NkT_{ni} \right)$ | Variable |
| Light intensity | $I = E^2 \Gamma \tau$ | Variable |
| Interaction between mesogens | $\hat{J} = J / \left( kT_{ni} \right)$ | 4.54 |
| Fraction of photoactive mesogens | $f$ | 1/6 |

## 3. Free energy landscape and stress-stretch responses

The photomechanical actuation being studied in this paper involves phase transformation with non-convex free energy landscape. To obtain key insights into the interplay between stress, light intensity, and temperature, we first investigate the landscape of free energy in Equation (1) under different conditions.

To start, we consider no mechanical stress ($\hat{s} = 0$) and examine the nematic-isotropic phase transformation in the LCE induced by either light illumination or increasing temperature. With no illumination ($I = 0$, **Fig. 2a**), the free energy landscape has a global minimum at about $Q = 0.6\text{-}0.7$ when the temperature is below the nematic-isotropic transition temperature $T_{ni}$, indicating the nematic phase with a finite degree of mesogen alignment. As the temperature increases to $\hat{T} = T / T_{ni} = 1$, a double-well free energy structure with two global minima (at $Q = 0$ and $Q = \sim0.4$) and an energy barrier between them emerges. The energy barrier is relatively small and hardly observed in the plot, but is confirmed for its existence numerically, indicating the weak first-order nematic-isotropic phase transition. The further increase of temperature leads to the global energy minimum at $Q = 0$, corresponding to the stable isotropic phase when $T > T_{ni}$. For comparison, under an increasing light intensity with no mechanical stress ($\hat{s} = 0$) and a fixed temperature $\hat{T} = 0.91$ (which we approximate as the room temperature), the evolution of the free energy landscape (**Fig. 2b**) follows the same trend. As a result, in the current loading condition, the reduction of nematic order due to photochemistry can be considered as an effective temperature increase.

Several previous studies have shown that in a liquid crystal elastomer compared to a liquid crystal melt or solution, the nematic-isotropic phase transition becomes second-order or supercritical [28, 38-41]. It has been hypothesized that such a change from first-order to second-order results from the introduction of polymer chains, crosslinking, and the internal residual stress during the material synthesis. These effects are not included in the current paper, and are currently being investigated with detail in a follow-up work.

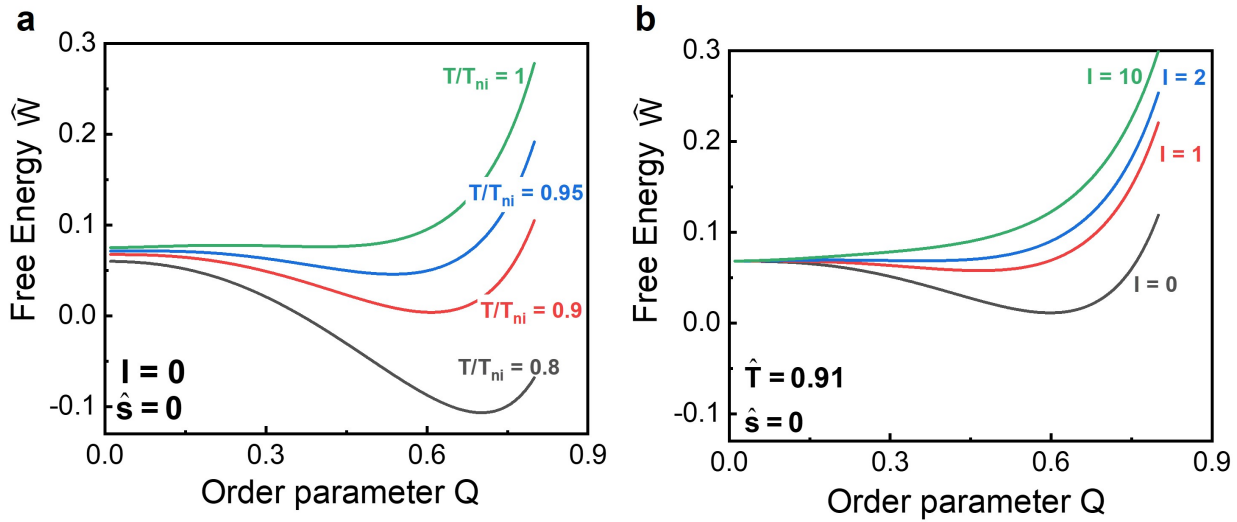

**Fig. 2.** The free energy landscape in the LCE with no mechanical stress ($\hat{s} = 0$) as a function of the order parameter $Q$ under: **(a)** no light illumination $I = 0$ and different temperatures $\hat{T} = T / T_{ni}$; **(b)** fixed temperature $\hat{T} = 0.91$ and increasing light intensity $I$.

This temperature- or light-induced nematic-isotropic transition further depends on the state of mechanical stress in the LCE. As indicated in both **Fig. 3a** (no light with different temperatures) and **Fig. 3b** (fixed temperature with different light intensities), a finite tensile stress from $\hat{s} = 0$ to $\hat{s} = 2$ effectively increases the nematic-isotropic transition temperature above $T_{ni}$. To illustrate, when a stress-free LCE is in the nematic phase (e.g., solid black curves with $\hat{s} = 0$ in Fig. 3a&b), the increased stress (dashed black curves with $\hat{s} = 2$) shifts the global free energy minimum towards a larger $Q$. Furthermore, when the stress-free LCE is in the isotropic phase (e.g., solid green curves with $\hat{s} = 0$ in Fig. 3a&b), the increased stress (dashed green curves with $\hat{s} = 2$) tilts the energy landscape with its minimum from $Q = 0$ to a finite $Q$, transforming the LCE back to the nematic phase.

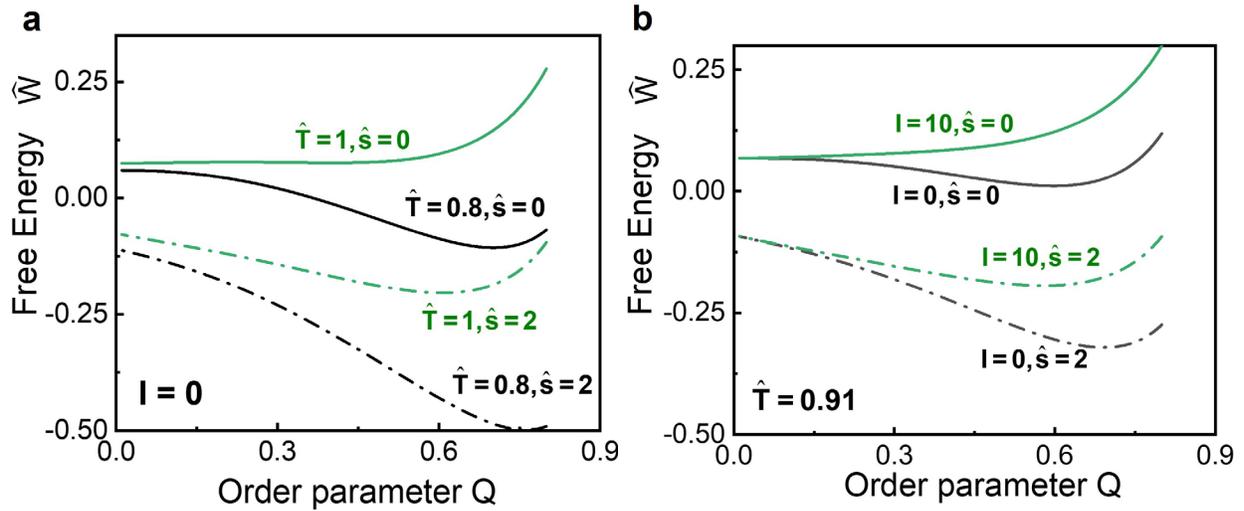

**Fig. 3.** The effect of mechanical stress on the photomechanical free energy landscape. An increase of stress from $\hat{s} = 0$ to $\hat{s} = 2$ effectively increases the nematic-isotropic transition temperature. **(a)** $I = 0$ with different $\hat{T}$ and $\hat{s}$. **(b)** $\hat{T} = 0.91$ with different $I$ and $\hat{s}$.

Different temperatures and light intensities also affect the uniaxial stress-stretch responses of the LCE. The constitutive model from minimizing the free energy in Equation (1) is simplified as

$$s = NkT \left( \frac{\lambda}{1 + 2Q} - \frac{1}{(1 - Q)\lambda^2} \right). \tag{7}$$

For each prescribed stress $\hat{s}$, we search the global free energy minimum to find the equilibrium stretch $\lambda$. We plot the stress-stretch curves with different temperatures and no light illumination ($I = 0$) in **Fig. 4a**, and with different light intensities and a fixed temperature ($\hat{T} = 0.91$) in **Fig. 4b**.

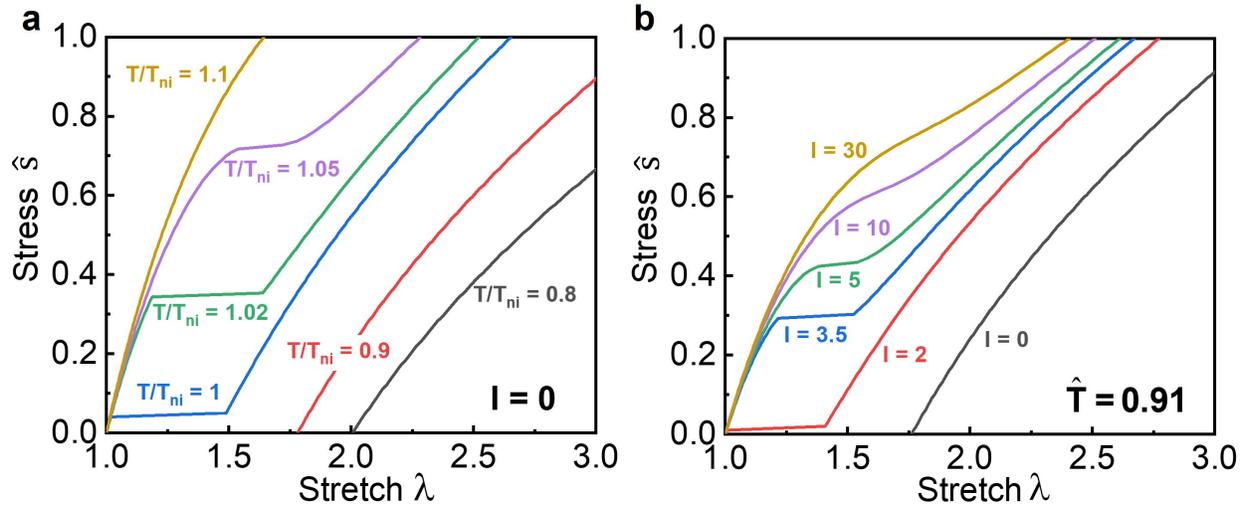

**Fig. 4.** Uniaxial stress-stretch responses of the LCE under different temperatures and light intensities. **(a)** $I$ = 0 with different $\hat{T} = T / T_{ni}$. **(b)** $\hat{T} = 0.91$ with different $I$.

Because the high-temperature isotropic state is chosen as the reference state, the LCE in Fig. 4a shows a finite stretch $\lambda > 1$ at zero stress when the temperature is below $T_{ni}$, due to the spontaneous deformation from the isotropic-nematic phase transition. As the temperature increases to above $T_{ni}$, the LCE transforms to the isotropic phase, and the stress-free spontaneous deformation vanishes. The stress-stretch curve shows a near-flat plateau with temperature close to and above $T_{ni}$ (e.g., $\hat{T} = 1$ and 1.02 in Fig. 4a), indicating the stress-induced transformation from the isotropic phase back to the nematic phase. The discontinuity in the slope results from the first-order nature of the nematic-isotropic transformation. When the temperature $\hat{T}$ becomes large enough, the stress-stretch curve becomes smooth again, the first-order nature disappears, and a supercritical second-order transformation emerges. An LCE under light illumination with fixed temperature shows similar stress-stretch behaviors (Fig. 4b) with light-induced nematic-isotropic phase transformation, stress-induced isotropic-nematic back transformation, and their first-to-second-order evolution with increasing light intensity.

The theoretical results in Fig. 4a agree remarkably well with previous experimental and modeling results in the literature [38, 42], while the results in Fig. 4b call for further experimental investigation. The above results reconfirm a long-existing hypothesis since Finkelmann et al. [8] that has been adapted by most existing theoretical models for photoactive LCEs [24, 36, 43, 44]: the light-induced isomerization can be considered as a temperature increase as a function of light intensity, such that the photomechanical actuation is modeled as an effective thermomechanical response of the LCE. This analogy is only valid when the actuation does not involve a reorientation of the nematic director, as in the case studied here. When director reorientation occurs in a photomechanical actuation (e.g., induced by polarized light), the current theory based on the physical entropic elastic model and the Maier-Saupe free energy is capable of capturing new phenomena involving the inherent coupling between stress, light, nematic director, and deformation [27]. The theoretical results above also predict a stress-induced increase of $T_{ni}$, which has been neglected in most studies, but has been both observed in experiment (by a few Kelvins) and analyzed by the Landau–de Gennes theory [38, 45].

## 4. Temperature-modulated snap-through, specific work, and blocking stress

We next use the theoretical framework to exploit the coupling between individual controls of temperature and light in a single photomechanical actuation, to enable a more versatile tuning of the photomechanical response compared to utilizing only light or temperature. This will be explored through the temperature-modulated snap-through instability, specific work, and blocking stress in a photomechanical LCE actuator.

Our preceding work has shown that the light-induced nematic-isotropic phase transformation may lead to a snap-through instability in the LCE [27]. Here we further investigate the temperature modulation of this snap-through behavior. We consider the same sheet of LCE as in Fig. 1b subjected to a constant stress $\hat{s}$ and increasing light intensity, at different temperatures. With a relatively small stress ($\hat{s} = 0.25$, **Fig. 5a**), the snap-through is not observed within the range of light intensity being investigated when the temperature $\hat{T} = T / T_{ni}$ is low ($\hat{T} = 0.85$), but is observed when $\hat{T}$ increases ($\hat{T} = 0.9$). The LCE is in the nematic phase with a finite stretch at $I = 0$, and gradually contracts with the increasing intensity, until the curve reaches a peak. The further increase of $I$ beyond this peak causes an instant snap to a much smaller stretch, indicating the light-induced nematic-isotropic transformation. Similarly, the decrease of $I$ from a large value leads to a backward snap-through response.

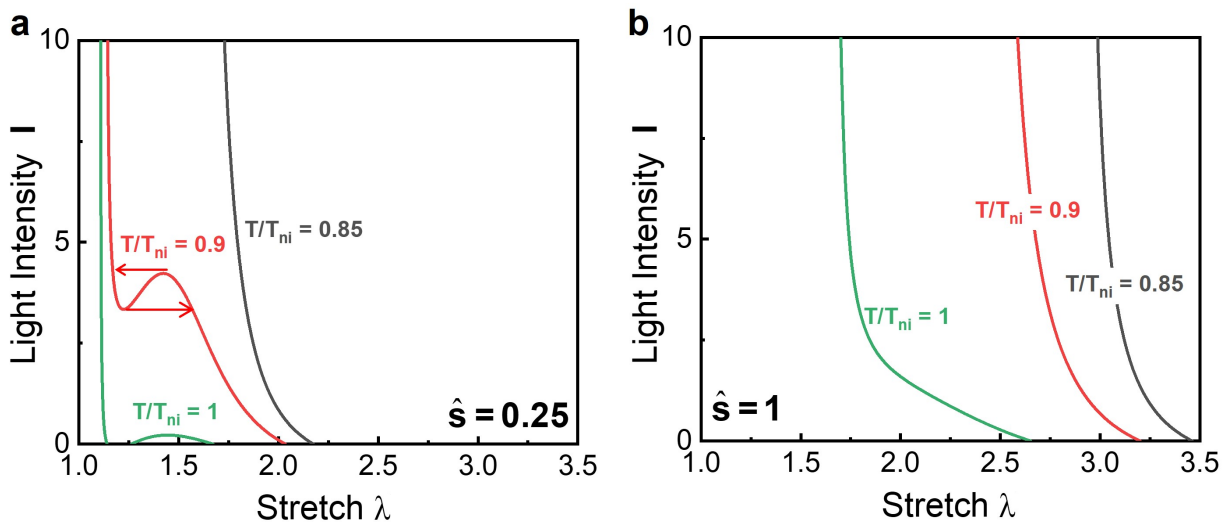

**Fig. 5.** Temperature-modulated photomechanical snap-through instability. The dimensionless light intensity $I$ is plotted as a function of the stretch $\lambda$ with different temperatures and stresses. **(a)** $\hat{s} = 0.25$. **(b)** $\hat{s} = 1$.

This temperature-modulated snap-through can again be well explained by the effective temperature increase from the light illumination. When the temperature is relatively low ($\hat{T} = 0.85$), the range of light intensity investigated in Fig. 5 is not large enough for an effective temperature increase to reach the nematic-isotropic transformation, thus no snap-through occurs. An increase of temperature to $\hat{T} = 0.9$ makes it closer to the transition temperature $\hat{T} = 1$, such that snap-through becomes available at a lower light intensity.

Of further particular interest is the case of $\hat{T} = 1$ and $\hat{s} = 0.25$ (green curve) in Fig. 5a. A similar snap-through is observed, where the curve has a part below $I = 0$ with a seemingly "negative light intensity". This negative light intensity reflects an effective decrease of temperature. The curve qualitatively agrees very well with the previous theoretical modeling of thermomechanical LCEs using the entropic elastic model and the Landau–de Gennes theory [45].

This observation of negative light intensity in Fig. 5a indicates a potential new way of tuning the photomechanical actuation using individual controls of light and temperature. At $\hat{T} = 1$ and $\hat{s} = 0.25$, the LCE initially stays in the nematic phase very close to isotropic without light due to the nonzero tensile stress $\hat{s} = 0.25$ which effectively increases $T_{ni}$. During the actuation, since it is impossible to induce a negative light intensity in real working scenario, the decrease of light intensity from a large value to $I = 0$ will end up being insufficient to induce the backward snap-through. The LCE is subsequently trapped in the metastable isotropic phase with a stretch ($\sim 1.2$) much smaller than its spontaneous nematic stretch.

When the stress is large enough ($\hat{s} = 1$, **Fig. 5b**), it not only increases the transition temperature $T_{ni}$, but further suppresses any first-order phase transformation regardless of the varying temperature. This result is consistent with our previous study in the photomechanical phase diagram [27].

We then use the current setup to explore the maximum specific work that can be generated by an LCE actuator under illumination of a constant light intensity and variable temperature. **Fig. 6** illustrates such an actuation process. An LCE sheet with a fixed load denoted by $\hat{s}$ and a prescribed temperature $\hat{T}$ contracts by a stretch $\delta\lambda$ under the light illumination of constant intensity $I$. The *specific work* generated by this photomechanical actuation is $\hat{s}\delta\lambda$, defined as the generated work per unit volume of the LCE sheet. We are interested in optimizing $\delta\lambda$ (thus the specific work $\hat{s}\delta\lambda$ under a fixed load $\hat{s}$) over the material temperature $\hat{T}$.

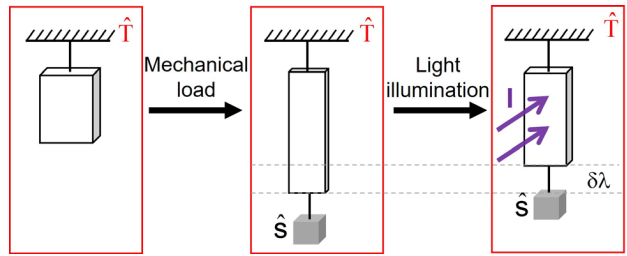

**Fig. 6.** An LCE sheet with a fixed load denoted by $\hat{s}$ and a prescribed temperature $\hat{T}$ contracts by a stretch $\delta\lambda$ under the light illumination. The specific work generated by this photomechanical actuation is $\hat{s}\delta\lambda$.

We study three cases of different mechanical stresses in **Fig. 7**: $\hat{s} = 0$, 0.5 and 1.5. The stretch change $\delta\lambda$ reaches a peak at an intermediate temperature in the first two cases ($\hat{s} = 0$ and 0.5, Fig. 7a&b) but remains monotonic with $\hat{T}$ in the third case ($\hat{s} = 1.5$, Fig. 7c). This intriguing behavior can again be explained by the light-induced effective temperature increase. When the stress is small as in Fig. 7a&b, at a low temperature before the peak, the finite constant light illumination is not strong enough to trigger the nematic-isotropic phase transformation. Further increasing the temperature leads to a rather steep increase of $\delta\lambda$, as a direct result of the nematic-isotropic transformation. $\delta\lambda$ then decreases with the increasing $\hat{T}$ after the peak, since the material becomes closer to the isotropic phase without illumination such that the transformation-induced photomechanical actuation becomes less effective.

The stretch change $\delta\lambda$ also depends on the fixed light intensity $I$ and stress $\hat{s}$. As shown in Fig. 7a&b, a higher $I$ shifts the peak towards a smaller $\hat{T}$ with a larger peak value of $\delta\lambda$. As the stress $\hat{s}$ increases, the curves show a clear transition from the first-order (Fig. 7a) to second-order (Fig. 7b) phase transformation, and finally no transformation with a monotonic curve throughout the temperature range being examined (Fig. 7c). All these observations can be similarly explained by the light-induced effective temperature increase and the stress-induced suppression of nematic-isotropic phase transformation.

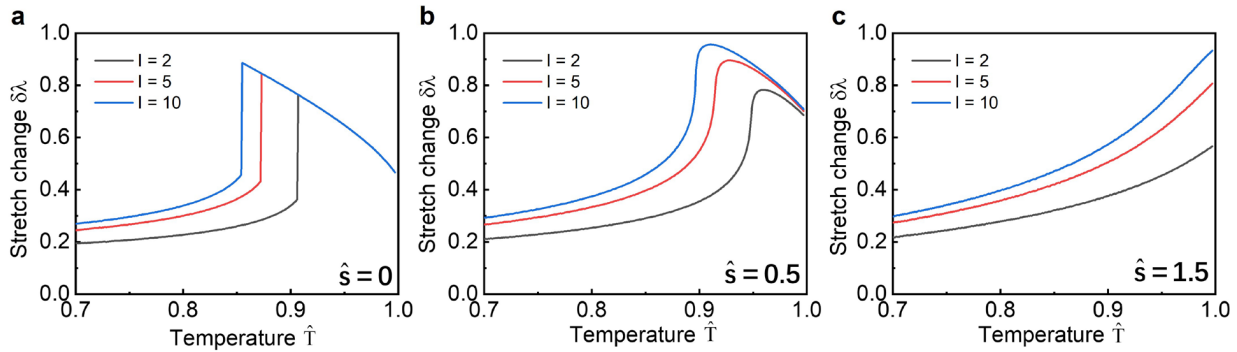

**Fig. 7.** The stretch change $\delta\lambda$ as a function of the temperature $\hat{T}$ for an LCE sheet subjected to a fixed load $\hat{s}$ and light intensity $I$. **(a)** $\hat{s} = 0$. **(b)** $\hat{s} = 0.5$. **(c)** $\hat{s} = 1.5$.

Finally, we use the model to examine the blocking stress of a photomechanical transducer that was proposed by Cviklinski, Tajbakhsh, and Terentjev in one of the first few papers on photoactive LCEs [32]. As illustrated in **Fig. 8a**, a nematic LCE sheet initially stays in the stress-free state with a spontaneous deformation under a prescribed temperature $\hat{T}$. The sheet is clamped by both ends and illuminated by light of intensity $I$. Since the length of sheet is fixed during the process, a tensile stress change $\delta\hat{s}$ builds up in the sheet due to the light illumination. We call this tensile stress change the *blocking stress*, just like the *blocking force* as the maximum force produced by an actuator when its movement is blocked.

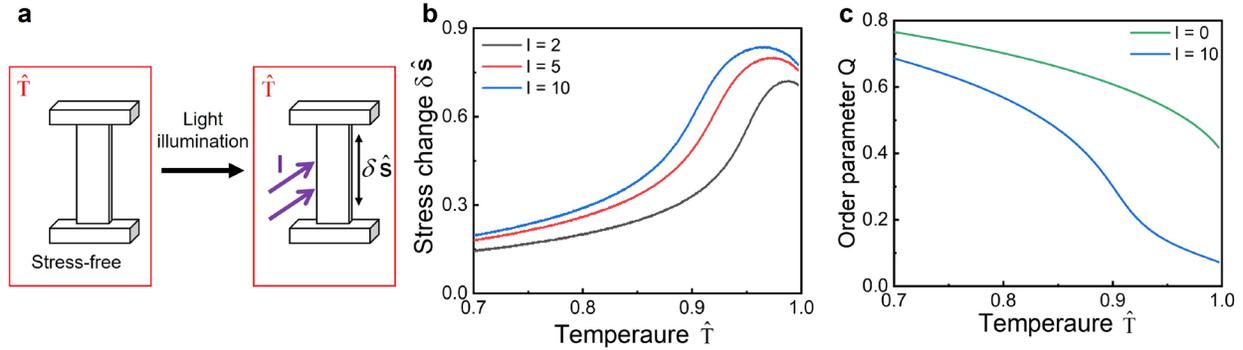

**Fig. 8.** **(a)** An LCE sheet initially stays in the stress-free state with a spontaneous deformation under a prescribed temperature $\hat{T}$. The sheet is clamped by both ends and illuminated by light of intensity $I$. Since the length of sheet is fixed during the process, a tensile stress change $\delta\hat{s}$ builds up in the sheet due to the light illumination. **(b)** The stress change $\delta\hat{s}$ as a function of the temperature $\hat{T}$ with different light intensities $I$. **(c)** The order parameter as a function of temperature $\hat{T}$ prior to ($I = 0$) and under the light illumination ($I = 10$).

**Fig. 8b** plots the blocking stress $\delta\hat{s}$ as a function of the prescribed temperature $\hat{T}$ with different light intensities. When $\hat{T} = 1$ and above, all the curves collapse to $\delta\hat{s} = 0$ (not plotted in the figure) since the light illumination does not influence a stress-free LCE in the isotropic phase. Below $\hat{T} = 1$, a nonmonotonic behavior shows up with a peak at an intermediate temperature. A higher light intensity $I$ increases the peak value and shifts the peak towards a smaller $\hat{T}$. These phenomena are consistent with what have been observed in Fig. 7, since the two working scenarios in Fig. 6 and Fig. 8a are inherently connected, analogous to the connection between the thermal strain and thermal stress in a linear elastic material. This nonmonotonic stress change with temperature qualitatively agrees very well with the experimental results reported before [32].

Interestingly, in the current scenario, the LCE always stays in the nematic phase without any phase transformation throughout the temperature range examined ($\hat{T} = 0.7$-1). This is examined by the calculated nematic order parameter $Q$ prior to ($I = 0$) and under the light illumination ($I = 10$) in **Fig. 8c**. The order parameter stays between 0 and 1 all the time and decreases monotonically with the increasing temperature $\hat{T}$. This marked difference in the temperature dependence between the blocking stress (nonmonotonic) and the order parameter (monotonic) highlights the profound interplay between the mechanical loading condition, the photoreaction, and the liquid crystal nematic order.

## 5. Temperature effect through thermal relaxation of chromophores

In the above discussions, the effect of temperature through the kinetics of thermal-induced *cis-trans* backward isomerization has been neglected. We now investigate this effect using the classical Arrhenius equation [46, 47]. Recall the dimensionless light intensity defined previously, $I = E^2 \Gamma \tau$, where $\tau$ is the thermal-induced relaxation time of the *cis* chromophore back to *trans*. The reciprocal of $\tau$ represents the reaction kinetics, which is assumed to follow the Arrhenius equation as

$$\frac{1}{\tau} \sim \exp\left(-\frac{E_a}{kT}\right), \tag{8}$$

where $E_a$ is the activation energy. Since we are only interested in the qualitative behavior in the current model, Equation (9) is simplified as

$$\tau = \tau_0 \exp\left(\frac{\hat{E}}{\hat{T}}\right), \tag{9}$$

where $\tau_0$ is a reference relaxation time and $\hat{E} = E_a / (kT_{ni})$ is the dimensionless activation energy. To better analyze the temperature effect, we further introduce a modified dimensionless light intensity, $\hat{I} = E^2 \Gamma \tau_0$, to make it independent of $\hat{T}$. The relationship between $I$ and $\hat{I}$ is thus

$$I = \hat{I} \exp\left(\frac{\hat{E}}{\hat{T}}\right). \tag{10}$$

When the thermal relaxation is assumed to be temperature-independent, as in the preceding discussions, $\hat{E}$ is set to zero and $I = \hat{I}$. By substituting Equation (9) into (6), we incorporate the temperature-dependent thermal relaxation in the model.

We use the new model to repeat the working scenario in Fig. 6, and plot similar curves in **Fig. 9** as those in Fig. 7. The results without considering the temperature-dependent thermal relaxation are plotted as $\hat{E} = 0$. The dimensionless light intensity is fixed as $\hat{I} = 2$. In all the cases with different stress $\hat{s}$ in Fig. 9, increasing the dimensionless activation energy $\hat{E}$ shifts the curve towards a lower temperature, equivalent to increasing the light intensity in the system as previously shown in Fig. 7. Indeed, a larger $\hat{E}$ corresponds to a higher energy barrier and thus more suppresses the thermal-induced *cis-trans* backward reaction.

The fact that the two trends with increasing $\hat{E}$ in Fig. 9 and increasing $I$ in Fig. 7 are nearly identical alludes to the intrinsic competition between two temperature-dependent processes: the *cis-trans* thermal relaxation and the nematic-isotropic phase transformation. While the former is mainly a process of individual molecules, the latter involves a collective process of many molecules with long-range directional interactions. Since the curves in Fig. 9 do not significantly change from Fig. 7 which only considers the temperature-dependent nematic-isotropic transformation, we conclude that the collective process from long-range molecular interactions (that results in the nematic-isotropic transformation) plays a dominant effect over the single-molecule thermal relaxation, in the temperature dependence of photomechanical actuation investigated here. This conclusion is consistent with our recent studies on solid-state photoreactions in other photomechanical systems [10, 11], and calls for future experimental validation.

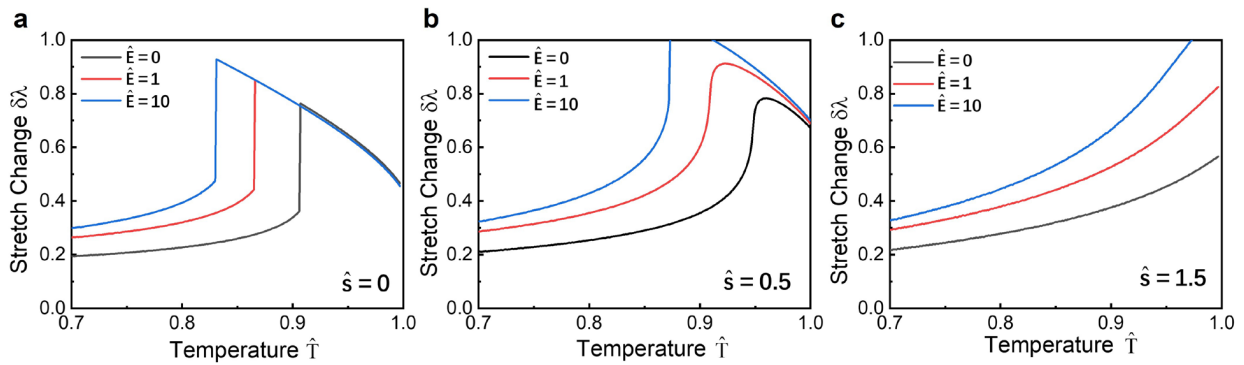

**Fig. 9.** The temperature effect through thermal relaxation of chromophores. The stretch change $\delta\lambda$ is plotted as a function of the temperature $\hat{T}$ for an LCE sheet subjected to a fixed light intensity $\hat{I} = 2$ and prescribed stress: **(a)** $\hat{s} = 0$; **(b)** $\hat{s} = 0.5$; **(c)** $\hat{s} = 1.5$. The activation energy is selected as $\hat{E} = 0$, 1, and 10 in each figure.

## 6. Discussion and conclusion

The temperature effect on mechanical behaviors of LCEs has been extensively investigated predating the photochemical effect and is still actively studied nowadays. Temperature itself is a common and useful way of actuating LCEs. Photothermal-induced actuations of LCEs have become particularly attractive in recent years with the advancements in both the fundamental understanding of the geometry-kinetics coupling and new applications in soft robotics and shape morphing [48-51]. In the current paper, we have demonstrated several working scenarios with interesting photomechanical responses of an LCE under individual controls of light and temperature. These theoretical demonstrations open the door to a new design concept, that one can combine both photochemical and photothermal effects in a single light illumination process, analogous to the recent idea of using dual-wavelength lights for better photochemical actuation [44]. As an example, one can embed light-absorbing nanoparticles into a photoactive LCE, and subsequently shine two light beams of different wavelengths to the material. One light beam (such as UV) triggers the photochemistry, while the other light (NIR) heats up the nanoparticles to control the temperature. This separate control will thus enable the optimization of different LCE responses (such as the work output and blocking stress discussed in this paper) for specific applications.

The key experimental challenge to achieve photomechanical responses reported in this paper is the relatively small light penetration depth into the LCE. A high concentration of azobenzene promotes the local photoreaction, but the reacted *cis* azobenzene limits the light to further propagate into the material. To achieve a light-induced homogeneous deformation through thickness, previous experimental results suggest that a thickness on the order of 0.4 mm and azobenzene concentration about 30 mol% are possible [32, 36]. The condition of homogeneous deformation can be directly examined by observation of any possible bending under illumination, or by measuring the light penetration depth using absorption [7].

The current model does not consider the reorientation of nematic director under the illumination of polarized light. This is often a valid assumption for a monodomain LCE synthesized in the nematic phase by common methods such as the widely used *two-stage polymerization* [37, 52], where a "memory" effect is built in the polymer chains by mechanical stretch [27].

To sum up, this paper builds a theoretical model of photoactive LCE to explore its temperature-modulated photomechanical actuation. The thermal- and photochemical-induced nematic-isotropic phase transformation in the LCE depends critically on the temperature, light intensity, and mechanical stress. We focus on a monodomain nematic LCE sheet with the tensile stress and light polarization both parallel with the nematic director. For the photomechanical response, the light illumination effectively increases the LCE temperature, while the tensile stress effectively increases the nematic-isotropic transition temperature and further suppresses the phase transformation when the stress is large. Several working scenarios have been theoretically investigated, where a combination of individual controls of light and temperature gives rise to interesting photomechanical responses as well as optimal performances. We hope these theoretical results will motivate further experimentation and help develop new actuation modes for nontraditional applications of photoactive LCEs. This paper is also hoped to shine light on future fundamental studies on the temperature-light-stress interplay in other photomechanical systems, such as polydomain photoactive LCEs with complex loading conditions.

**Acknowledgements**

This work was supported by the startup fund from Northeastern University.